\documentstyle [12pt,a4wide]{article}
\def\ftilde{f}
\def\htilde{h}
\def\calK{{\cal K}}

\def\diffsigma{{\rm Diff}\Sigma}

\def\ldiffsigma{{\rm LDiff}\Sigma}

\def\gwx{g^{\omega/2}}
\def\gwxp{g^{'{\omega/2}}}

\def\ham{{\cal H}_{\perp}}

\def\mom{{\cal H}_i }
\def\momj{{\cal H}_j }

\def\phiprtwo{{1\over2}\phi '(\alpha)}

\def\xxp{(x\leftrightarrow x')}

\begin{document} 
 
\thispagestyle{empty}
\setcounter{page}{1}

\begin{flushright}
{\rm IMPERIAL/TP/95--96/20}
\end{flushright}
\vskip 1cm
\begin{center}
{\large{\bf Gravitational Constraint Combinations\\
Generate a Lie Algebra.}}
\vskip 2cm
{F.G.~Markopoulou\footnote{email: f.markopoulou@ic.ac.uk}
\vskip 0.4cm
{\it Theoretical Physics Group\\
Blackett Laboratory\\ 
Imperial College of Science, Technology \& Medicine\\
London SW7 2BZ, U.K.}}
\end{center}


\begin{abstract}
We find a first--order partial differential equation whose
solutions are all ultralocal scalar combinations of gravitational 
constraints with  Abelian Poisson brackets between themselves. 
This is a generalisation of the Kucha\v{r} idea of finding 
alternative constraints for canonical
gravity.  The new scalars may be used
in place of the hamiltonian constraint of general relativity and,
together with the usual momentum constraints, 
replace the Dirac algebra for pure gravity with a true Lie algebra:
the semidirect product of the Abelian algebra of the new constraint
combinations with the algebra of spatial diffeomorphisms.  
\end{abstract}

\small{PACS} numbers: 0420, 0460.

\newpage

\section{Introduction}
In the hamiltonian formulation of general relativity one assumes that
the spacetime manifold $\cal M$, with  metric
$\gamma_{\alpha\beta}(X)$,  can be foliated as $\Sigma\times R$ where
$\Sigma$ is a three--dimensional manifold and the real line provides a
global time coordinate.\footnote{
The notation used in this paper is: events on $M$
are denoted by $X^{\alpha}$, while spatial points on $\Sigma$ are 
labelled by $x^i$. Indices $\alpha,\beta,...$ are spacetime indices
running from 0 to 3; $i,j,...$ are indices on $\Sigma$ that run from 1
to 3. A standard notation is used to distinguish  functions from 
functionals, namely, round brackets, ( ), indicate that the object is
a function of the 
arguments in the brackets, square brackets, [ ], denote a functional and
an object with a  mixed bracket,
( ; ], is a function of the arguments that appear on the
left and a functional of those on the right. Finally, the covariant
derivative on $(\Sigma, g)$ is written as $D_i$.}
The physical information of the four--dimensional theory is
then contained in the intrinsic metric $g_{ij}(x)$ on $\Sigma$---the
pullback of  $\gamma_{\alpha\beta}(X)$ on $\Sigma$---and its conjugate
momentum $p^{ij}(x)$. 

On the foliated manifold, the Einstein equations of general
relativity become ten equations that the canonical 
data $(g_{ij}(x),p^{ij}(x))$ must satisfy.
Six of these are first--order dynamical equations and describe how 
the spatial metric and its conjugate momentum change with time;
the remaining  
four are constraints that $(g_{ij}(x),p^{ij}(x))$ must obey at any point
$x$ on $\Sigma$ and reflect the symmetries of the four--dimensional theory.  
The momentum constraints $\mom(x)$
 generate spatial diffeomorphisms on a slice $\Sigma$, and the
hamiltonian constraint $\ham(x)$  propagates 
$\Sigma$ in the direction normal to the hypersurface. 
As functions of $g_{ij}(x)$, $p^{ij}(x)$ and the
determinant of the spatial metric, $g(x)$, the 
gravitational constraints are\cite{{ADM},{Dirac}}:
\begin{eqnarray}
\ham(x;g,p]&=&G_{ijkl}(x;g]p^{ij}(x)p^{kl}(x) -g^{1\over2}(x)\
^{3}R(x;g]\\
 {\rm where} & &  G_{ijkl}(x;g]={1\over2}g^{-{1\over2}}(x)(
g_{ik}g_{jl}+g_{il}g_{jk}-g_{ij}g_{kl})\nonumber\\
\mom(x;g,p]&=&-2D_jp^j_i(x).
\label{eq:constraint}
\end{eqnarray}

The constraint system, 
\begin{equation}
\ham(x)=0=\mom(x),
\label{eq:con}
\end{equation}
satisfies the Dirac algebra
\begin{eqnarray}
\{\ham (x), \ham (x')\}&=&g^{ij}(x)\mom(x) \delta_{,j}(x,x')-\xxp
\label{eq:D1}
\\
\{\ham(x), \mom(x')\}&=&{\ham}_{,i}(x) \delta (x,x')+
\ham(x) \delta_{,i}(x,x')\label{eq:D2}
\\
\{\mom(x),\momj (x')\}&=&\momj(x)\delta_{,i}(x,x')-
(ix\leftrightarrow jx').
\label{eq:D3}
\end {eqnarray}
The Dirac algebra is often thought of as the algebra of
spacetime diffeomorphisms, ${\rm LDiff}M$,
``projected'' on the foliation  $\Sigma\times R$. 
Equation ({\ref{eq:D3}) is the algebra of spatial diffeomorphisms,
generated by $\mom(x)$. The Poisson bracket (\ref{eq:D2}) shows
that $\ham(x)$
transforms  as a scalar density of weight 1 under spatial
diffeomorphisms. The first equation, 
(\ref{eq:D1}), however,  means that two infinitesimal normal deformations
on $\Sigma$, performed in arbitrary order, end on the same final
hypersurface but 
not on the same point on that hypersurface\cite{KuDGR}. 
The fact that the right hand side of this Poisson bracket involves the
metric explicitly is a source of problems
in any attempt to use the Dirac algebra in a canonical quantisation of
gravity. The Dirac 
algebra is closed, in the sense that the  Poisson brackets between
constraints are constraints themselves, however, the structure
coefficients in equation (\ref{eq:D1}) depend on the canonical
variables. 
In a quantum version of the theory where, presumably, quantum counterparts
of the classical gravitational data and the constraints will appear,
the Poisson bracket algebra will go over to a commutator algebra which 
is not a Lie algebra.  

Recently, Brown and Kucha\v{r}\cite{BrKu} obtained  some surprising
results which provide a promising proposal on the problem of the
algebra of canonical gravity. They studied a spacetime filled with
incoherent dust; in such a system the coupling of the metric to 
matter introduces into spacetime a privileged dynamical frame and time
foliation. The system has a very interesting feature:
 the scalar constraint for the combined system, which in \cite{BrKu}
was denoted by  $H_{\uparrow}(x)$,  can be
split into two separate parts, one for
matter and one for gravity, and cast in the form
\begin{equation}
H_\uparrow(x) := P(x) + h(x; \, g_{ij},p^{ij}] = 0, 
\end {equation}
where $P(x)$ is the momentum conjugate to the ``dust time'' variable
$T(x)$, and 
$h(x)$ is the following scalar combination of the gravitational
constraints:
\begin{equation}
h(x)^2 =G(x):=\ham ^2(x)-g^{ij}\mom(x) \momj(x).
\label{eq:G}
\end {equation}

The truly remarkable result of \cite{BrKu} was that $G(x)$ and $h(x)$,
scalar functions of gravitational variables only, have strongly
vanishing Poisson brackets among themselves. One then has a
possible alternative set of constraints for pure gravity,
\begin{equation}
G(x)=0=\mom(x),
\end{equation}
 which generate the true Lie algebra,
\begin{eqnarray}
\{G(x),G(x')\}&=&0\,,\label{eq:GG} \\
\{G(x), \mom (x')\}&=&G_{,i}(x)\delta (x,x')+2G(x)\delta_{,i}(x,x')\,,
\label{eq:GH}\\
\{\mom (x), \momj (x')\}&=&\momj(x)\delta_{,i}(x,x')-
(ix\leftrightarrow jx'). \label{eq:HH} 
\end {eqnarray} 
This algebra  is the semidirect product of the Abelian algebra 
generated by $G(x)$  (equation \ref{eq:GG}), and the algebra of 
spatial diffeomorphisms 
$\ldiffsigma$ generated by $\mom(x)$ (equation \ref{eq:HH}). 
The Poisson bracket (\ref{eq:GH}) is the transformation 
of $G(x)$ as a weight 2 scalar density under $\diffsigma$.

Previous
investigations on the use of matter time and reference fluids in
canonical gravity\cite{Kuchar, Brown} led to the
``phenomenological'' approach to
the problem of time in canonical gravity of \cite{BrKu}. The
significance of the new 
constraints (\ref{eq:G}), however, and the role they could play in
pure gravity was left to be investigated at some later stage.
That dust was not, in itself, essential in the process was made
clear when Kucha\v{r} and Romano\cite{KuRo} found a simpler way of
producing results similar to those of ref.~\cite{BrKu}. They coupled  
gravity to a massless scalar field, and the scalar density they
obtained is the following combination of gravitational constraints 
\begin{equation}
\Lambda_{\pm}(x)=
g^{1\over2}(x)\biggl(-\ham (x)\pm\sqrt{G(x)}\biggr). \label{eq:L}
\end {equation} 
As in the case of $G(x)$, the weight--two scalar densities 
$\Lambda_{\pm}(x)$ have strongly vanishing Poisson brackets with
themselves. They can be used to rewrite the constraint system for pure
gravity as
\begin{equation} 
\Lambda_{\pm}(x)=0=\mom (x) 
\end {equation} 
and, again, if $\Lambda_{\pm}(x)$ can be taken to replace $\ham(x)$ a
true Lie algebra for vacuum gravity has been found.  

Although coupling to matter was an essential step for \cite{BrKu} and 
\cite{KuRo} to find the explicit form of  the  new scalar
constraint, the resulting combinations, $G(x)$ and $\Lambda(x)$, involve
gravitational variables only and do not depend on the matter variables
to which the gravitational field was originally coupled. The question
that naturally arises therefore is whether a transformation of the scalar
constraint to a form that generates a true Lie algebra can be made
within the context of pure gravity only.

In this paper we take the first step towards answering this question
by finding the full set of scalar densities, combinations of the
original pure gravity constraints, that have the crucial property of
generating a Lie algebra (with vanishing Poisson brackets between
themselves). We find 
that the calculation can be carried 
out by  working directly with the Dirac algebra, without any reference
to matter couplings. Thus, we are not restricted by any need to
include matter fields and there is greater freedom in using the new
algebra in pure gravity. A complementary view of these combinations
of pure gravity constraints is provided in a paper by
Kouletsis\cite{iannis}, to appear shortly. In this work the
constraint combinations 
are shown to arise naturally in a system of pure gravity with 
non--derivative coupling to an action of a single
scalar field and two arbitrary functions of a Lagrange multiplier.

\section{Constraint combinations are generated by a \\
differential equation. }

From  the work of \cite{BrKu} and \cite{KuRo}, it seems clear that
there might be 
more combinations of gravitational constraints obeying the algebra
(\ref{eq:GG}--\ref{eq:HH}). We want to find  them  all, 
and furthermore, as explained in the introduction,  we wish to
investigate the new algebra without any 
reference to a matter coupling. We therefore choose to work directly
with the Dirac algebra of pure gravity. 

A particularly  interesting feature of the constraints $G(x)$ and
$\Lambda_{\pm}(x)$ is that 
they are  scalar densities of weight 2, unlike the usual hamiltonian
constraint $\ham(x)$ which is a weight 1 scalar density. An important
question therefore is whether a weight of 2 is in some way a  natural
choice for a scalar constraint for pure gravity
that has Abelian Poisson brackets with itself. 

For generality then, we assume that the candidate new
constraint we are looking for is an ultralocal  scalar
density $\calK(x)$ of arbitrary weight  $\omega$. Such a $\calK(x)$ can
only be a function of 
the scalars $\ham$, $(g^{ij}\mom\momj)$, and  the determinant of the
spatial metric $ g$ (there are no derivatives of the gravitational
constraints). 
Together with the usual momentum constraint,  $\calK(x)$ 
is assumed to satisfy the Lie algebra
\begin{eqnarray}
\{\calK (x), \calK (x')\}&=&0, \label{eq:WW}\\
\{\calK (x), \mom (x')\}&=&\calK_{,i}(x)\delta(x,x')+
\omega\calK(x)\delta_{,i}(x,x'),
 \label{eq:Wmom}\\
\{\mom(x),\momj (x')\}&=&\momj(x)\delta_{,i}(x,x')-
(ix\leftrightarrow jx').
\label{eq:mommom}
\end {eqnarray}
The bracket (\ref{eq:Wmom}), consistent with our original assumptions,
means simply that $\calK(x)$ transforms as  a scalar of weight
$\omega$ under $\diffsigma$.  The dependence of $\calK(x)$ on
$\ham(x)$ and $\mom(x)$ lies in the first bracket (\ref{eq:WW}), 
which we shall now evaluate.

The calculations that need to be done are  considerably simplified if we
define  $\htilde$ 
and $\ftilde$ as the two obvious scalar 
densities of weight zero that can be formed from the gravitational
constraints $\ham$ and $\mom$:
\begin{equation}
\htilde:=g^{-{1\over2}}\ham\qquad{\rm and}\qquad
\ftilde:=(g^{-1})g^{ij}\mom\momj.
\end{equation}
Using $\htilde$ and $\ftilde$, one can define another weight zero scalar
density $K(x)$, related to the weight $\omega$ density
$\calK(x)$ by 
\begin{equation}
\calK(x;\htilde,\ftilde,g]=g^{\omega/2}(x)K(x;\htilde,\ftilde].
\label{eq:twow}
\end {equation}
 It is easy to construct a $\calK(x)$ by premultiplying $K(x)$ by
the $g^{\omega/2}(x)$ factor, so for simplicity, our discussion is
centred around $K(x)$. One only needs to remember that 
$K(x)$ itself does not satisfy the Abelian Poisson bracket (\ref{eq:WW}). 
Its dependence on $\htilde$, $\ftilde$, however,  is the same as that
of $\calK(x)$. 

Imposing the central  requirement that  $\calK$ satisfies the
Poisson bracket (\ref{eq:WW}), we can now find, 
by simply working out the Poisson brackets,
which  (weight zero densities) $K$ 
correspond to the (weight $\omega$ densities) $\calK$ that generate a 
true Lie algebra .
Our calculation is a generalised analogue of the calculation of
Kucha\v{r} and Romano, so we concentrate on the steps needed to adapt
their calculation for arbitrary--weight densities. 

By the Leibniz rule, equation (\ref{eq:WW}) gives
\begin{equation}
\{\calK, \calK'\}=\gwx\gwxp\{K, K'\} +\biggl(\gwxp K \{\gwx, K'\}
-\xxp\biggr),
\label{eq:WWLR}
\end{equation}
where, to avoid cluttering of symbols, primes have been used on
quantities that have $x'$ as their spatial argument. The Poisson
brackets above can be expanded to
\begin{eqnarray}
\{K,K'\}&=&{\partial K\over\partial\htilde}
{\partial K'\over\partial\htilde'}
\{\htilde,\htilde'\}+
{\partial K\over\partial\ftilde}{\partial K'\over\partial\ftilde'}
\{\ftilde,\ftilde'\}\nonumber\\
& & +\biggl({\partial K\over\partial\htilde}
{\partial K'\over\partial\ftilde'}
\{\htilde,\ftilde'\}-\xxp\biggr),\\
 \{g^{\omega/ 2},K'\}&=&
{\partial K'\over\partial \htilde'}\{g^{\omega/2},\htilde'\}+
{\partial K'\over\partial \ftilde'}\{g^{\omega/2},\ftilde'\},
\label{eq:gW}
\end{eqnarray}
and therefore, one needs to calculate Poisson brackets 
 involving $\htilde$, $\ftilde$ and
$g^{\omega/2}$. 
To handle these  brackets,  the identity
\begin{equation}
\{\gwx, \ftilde'\}= \omega g ^{{\omega - 1\over 2}} \{g^{1\over2}, \ftilde'\},
\end {equation}
is used, 
while the bracket between $\gwx$ and $\htilde$ is proportional to
$\delta (x, x')$ and therefore its contribution cancels with the
$\xxp$ terms that appear in equation (\ref{eq:WWLR}). 
For the brackets of the weight zero gravitational variables $\htilde$
and $\ftilde$, it is straightforward,  using 
 the Jacobi identity and the standard Dirac
relations (\ref{eq:D1})--(\ref{eq:D3}),  to verify
that they satisfy the algebra
\begin{eqnarray}
 \{ \htilde , \htilde '\}& =&g^{-{1\over
2}}g^{'{-{1\over2}}}\{\ham,\ham '\} , \\
\{\ftilde ,\ftilde '\}&=&-4g^{-{1\over2}}g'^{-{1\over2}}\ftilde
\{\ham,\ham '\},\\
\{\htilde ,\ftilde'\}&=&\propto\delta(x,x').
\end{eqnarray}
where $ \propto \delta(x,x')$ stands for terms proportional to
$\delta(x,x')$ which, as always, cancel with the $\xxp$ terms.
One now has all the necessary information to extend the
procedure of Kucha\v{r} and Romano\cite{KuRo} for $\calK(x)$. 

The requirements on the scalar density $\calK(x)$
are minimal. Thus one would expect that the rather lengthy calculation
of the Poisson brackets between the $\calK$'s would lead to a
fairly cumbersome result. It is surprising to find it is instead quite
simple:
\begin{equation}
\{\calK , \calK '\}=g^{\omega-1}\Biggl [
\biggl({\partial K\over\partial\htilde}\biggr)^2
 - 4 \ftilde \biggl({\partial K\over\partial\ftilde}\biggr)^2 
+ 2 \omega K \biggl({\partial K\over\partial\ftilde}\biggr)
\Biggr]\bigl\{\ham,\ham'\bigr\}.
\label{eq:WWHH}
\end {equation}
All functions $K$ that turn the quantity inside the square brackets in
(\ref{eq:WWHH})   into
zero correspond, via equation (\ref{eq:twow}), to candidate new
constraints $\calK$. The family of these scalar functions are
therefore the
solutions of the nonlinear, first--order partial differential equation
for $K$
\begin{equation}
{ \omega \over 2} K\biggl({\partial K\over\partial\ftilde}\biggr) = 
\ftilde \biggl({\partial K\over\partial\ftilde}\biggr)^2 - {1\over 4}  
\biggl({\partial K\over\partial\htilde}\biggr)^2
\label{eq:WPDE}. 
\end {equation}

This unexpectedly compact equation is the key result 
because any of its solutions can be directly used to give a Lie
algebra for
pure gravity. Also, if by some other method one has already obtained a
combination of gravitational constraints which is suspected to satisfy
a Lie algebra, it is enough (and far easier) to instead check that it
satisfies the 
differential equation. For example, 
it can be easily verified that the differential equation
(\ref{eq:WPDE})  is satisfied by the scalars
that were discovered in references \cite{BrKu} and \cite{KuRo} (in their
respective weight zero form),
\begin{equation}
K\equiv g^{-1} G= \htilde^2 - \ftilde\qquad{\rm  and} 
\qquad K\equiv g^{-1}\Lambda_{\pm} = -\htilde \pm\sqrt {\htilde^2-\ftilde}, 
\end{equation}
for weight $\omega=2$. It is also satisfied by  the  weight $\omega=1$
scalar 
$\sqrt {\htilde^2-\ftilde}$ that 
appeared in ref. \cite{BrKu}.

\section{The solution of the differential equation}

The general solution to the first--order nonlinear partial differential
equation (\ref{eq:WPDE}) of the two independent variables $\htilde$ and 
$\ftilde$, for $\omega\neq 0$, is expected to be an expression
involving $\ham$, $\mom$ and an arbitrary function of one parameter.

A change of variables can turn (\ref{eq:WPDE}) into a more
manageable equation.  If we set
\begin{equation}
C=\ln K
\end{equation}
(we assume that $C$ in the above definition is well--defined)
and then divide the differential equation (\ref{eq:WPDE}) through by
$K^2$, we can rearrange and 
rewrite it as a partial differential equation for $C$:
\begin{equation}
\ftilde \Biggl(
{\partial C\over\partial \ftilde}\Biggr)^2-{\omega\over2}
{\partial C\over\partial \ftilde}-{1\over4}
\Biggl({\partial C\over\partial \htilde}\Biggr)^2=0.
\label{eq:KPDE}
\end{equation}
The advantage of this rearrangement is that this equation does not
involve $C$ explicitly. Because the variables $\htilde$, $\ftilde$ are
independent, the differential of the function $C(\htilde,\ftilde)$ 
 is given by the expression  
\begin{equation}
dC(\htilde,\ftilde)= \Biggl({\partial C\over\partial \ftilde}\Biggr) 
d\ftilde + \Biggl({\partial C\over\partial \htilde}\Biggr)d\htilde.
\label{eq:dK}
\end{equation}
In equation (\ref{eq:KPDE}) there are now no terms involving both
variables $\htilde$ and $\ftilde$, and thus it can be ``split'' into two
parts, for $\partial C/\partial\ftilde$ and $\partial C/\partial
\htilde$; namely, it holds \cite{pde} that any  solution to the coupled set 
\begin{eqnarray}
\ftilde \Biggl({\partial C\over\partial \ftilde}\Biggr)^2-
{\omega\over2}{\partial C\over\partial \ftilde}&=&\alpha^2\nonumber\\
{1\over4}\Biggl({\partial C\over\partial \htilde}\Biggr)^2&=&\alpha^2
\end{eqnarray}
for some constant $\alpha$, will also satisfy (\ref{eq:KPDE}). These 
two, single--variable,  equations can be readily solved as
\begin{equation}
{\partial C\over\partial \ftilde}={1\over4\ftilde}
\biggl(\omega\pm\sqrt{\omega^2+16\alpha^2\ftilde}\biggr) 
\qquad{\rm and}\qquad 
{\partial C\over\partial \htilde}=\pm 2\alpha
\end{equation}
(note that the $\pm$ signs appearing in the above pair of equations
are independent) and therefore the differential of $C(\htilde,\ftilde)$
in equation (\ref{eq:dK}) can be integrated to
\begin{equation}
C(\htilde,\ftilde,\alpha,\beta)={\omega\over4}\ln\ftilde\pm{1\over4}\Biggl[
2\sqrt{\omega^2+16\alpha^2\ftilde}+\omega\ln{
\sqrt{\omega^2+16\alpha^2\ftilde}-\omega\over
\sqrt{\omega^2+16\alpha^2\ftilde}+\omega}\Biggr]\pm
2\alpha\htilde+\beta
\label{eq:ci}
\end{equation}
(where, again, the two possible choices of $\pm$ sign are independent).
This is a complete integral of the differential equation for 
$C(\htilde,\ftilde)$;  namely,  a solution of the equation involving
two arbitrary independent constants $\alpha$, $\beta$ which come from
the two integrations in equation (\ref{eq:dK}). The complete integral
then is a two-parameter family of surfaces, since a different surface is
obtained for each choice of $\alpha$ and $\beta$. 

The general solution should also describe this family of surfaces. 
The difference is that, instead of depending on two arbitrary
parameters, the general solution 
should contain an arbitrary
function of one parameter. That is,  we suppose that 
\begin{equation}
\beta=\phi(\alpha)
\label{eq:beta}
\end{equation}
so that $C(\htilde,\ftilde,\alpha,\phi(\alpha))$ describes a
one-parameter family of solutions.  One can think of $\phi(\alpha)$ as
the  two-variable analogue of the integration
constant of the familiar one-variable  partial differential
equation.

One can go further by bringing in a basic theorem in the
theory of first--order partial 
differential 
equations \cite{pde}, that the envelope of any family of solutions of
a first order equation, depending on some parameter,
is again a solution. The envelope of the
one--parameter family of solutions $C(h,f,\alpha,\phi(\alpha))$ is
essentially a surface tangent to the family of 
surfaces that are the solutions. 
It should then be given by $C(\htilde,\ftilde,\alpha,\phi(\alpha))$,
together with its differential with respect to $\alpha$, namely,
\begin{equation}
{\partial C\over \partial\alpha}=
 \pm{\sqrt{\omega^2+16\alpha^2\ftilde}\over2\alpha}
\pm2\htilde+
\phi '(\alpha)=0,
\label{eq:envlp}
\end{equation}
where a prime denotes differentiation with respect to
$\alpha$.
One now only has to eliminate $\alpha$ from the above equation and
$C(\htilde,\ftilde,\alpha,\phi(\alpha))$ to have a single expression
for the envelope.  
This will also be the general solution involving
the arbitrary function $\phi(\alpha)$\cite{pde}. If we return to
the original equation for $K$ by exponentiating $C$ in equation
(\ref{eq:ci}), we can write out the general solution to the differential
equation (\ref{eq:WPDE}) directly. It is given by:
\begin{eqnarray}
K_{\pm}\bigl(\htilde, \ftilde, \phi(\alpha)\bigr)&=&
\Biggl[\biggl(\htilde \pm \phiprtwo \biggr) \pm\sqrt
{\biggl(\htilde \pm \phiprtwo \biggr)^2-\ftilde} \ \Biggr]
^{\omega\over 2}\nonumber\\
& & \times\exp\Biggl(\phi(\alpha)\mp{\omega\over2}{\phiprtwo\over 
\sqrt{\bigl(\htilde \pm\phiprtwo \bigr)^2-\ftilde}}\Biggr)
\label{eq:sln}
\end {eqnarray}
(where the $\pm$ signs in front of $\phi(\alpha)$ are independent of
those in front of the square roots)
together with equation (\ref{eq:envlp}). 
Note that, by a trivial rewriting of the exponential factor, the whole
solution can be raised to the power of  $(\omega/2)$, and hence the
constraint combination ${\cal 
K}(x)$, is simplest but non--trivial when $\omega=2$. This is
consistent with the fact that the ``physical''  matter couplings
used by Kucha\v{r} {\it et al.}, also produced weight 2 densities. 

Therefore,  for
each choice of the function $\phi(\alpha)$, one solves
(\ref{eq:envlp}) algebraically to obtain $\phi(\alpha)$ as a function of
$\htilde$ and $\ftilde$, and then substitutes the result in the
general solution (\ref{eq:sln}). For example, the Brown-Kucha\v{r}
solution is reproduced when $\phi(\alpha)=-\ln\alpha-1$ 
and $\omega=2$ (and
the minus signs in front of the square roots),
so that $\phi'(\alpha)=-{1\over\alpha}$. Then, solving
(\ref{eq:envlp}) gives  
\begin{equation}
\alpha=-{h\over h^2-f},
\end{equation}
which implies $\phi'(\alpha)=h-{f\over h}$, and 
which, when inserted in (\ref{eq:sln}), produces $K=\pm(h^2-f)$.
Obtaining the Kucha\v{r}-Romano solution is also straightforward, it
only requires setting $\phi(\alpha)=0$. 
With the same method one can also generate new solutions, in fact an
infinite number of them, for all possible choices of $\phi(\alpha)$.
New solutions, and also their relationship to scalar field
actions, have been investigated by Kouletsis in
\cite{iannis}.\footnote{
For example, the function $\phi(\alpha)=A\ln\alpha$, where $A$ is an arbitrary
constant, leads to the new solution\cite{iannis2}:
\begin{equation}
K=\biggl(-{h\over A}+R_1+R_2\biggr)\biggl({Ah+R_1\over 2(h^2-f)}\biggr)^A
\exp\biggl({Ah-R_1\over R_2}\biggr),
\end{equation}
where 
\begin{equation}
R_1=\sqrt{h^2+(A^2-1)f}\qquad{\rm and}\qquad R=\sqrt{{A^2+1\over A^2}
h^2+{A^2-1\over A^2}f-{2\over A}hR_1}.
\end{equation}
}
Generally, the only simple ones are those that have been
already found  by Kucha\v{r} {\it et al}.

The weight  $\omega=0$ case has to be treated separately since then
equation (\ref{eq:WPDE}) becomes a homogeneous 
partial differential equation,
\begin{equation}
 \ftilde \Biggl({\partial K^0\over \partial \ftilde}\Biggr)^2
 - {1\over 4}\Biggl({\partial K^0\over \partial \htilde}\Biggr)^2  = 0,
\end {equation}
whose solution is of the form $K^0 =\ \propto (\htilde 
\pm \sqrt{\ftilde})+{\rm constant}$. 

Summarising, we have shown that for each choice of the function
$\phi(\alpha)$ in the pair of equations (\ref{eq:envlp}) and
(\ref{eq:sln}), a combination of the gravitational constraints
$\htilde$ and $\ftilde$ is generated. The solution, via equation
(\ref{eq:twow}), provides the Abelian weight $\omega$ scalar density
${\calK}(x)$ which can be used in the construction of
a true Lie algebra for pure gravity.

\section{Conclusions}

In this paper we started from observations of the alternative
hamiltonian constraints found by Kucha\v{r} {\it et al.}, which
generate a true Lie algebra of gravity. We generalised to a much
larger family of such scalar densities in the hope that knowing their
general form would make their origin clearer.
All previous work on these constraints has been based on introducing some
simple form of matter into spacetime, either dust\cite{BrKu}, a scalar
field\cite{KuRo} or the deWitt action for a perfect fluid\cite{BrMa}.
We found that it is possible, and in fact simpler, to arrive at these
constraint combinations working solely with the algebra of
constraints. This is also the least restrictive method and enabled us
to generalise to ultralocal scalar densities of arbitrary weight.
Here we saw
that, although in principle scalar densities of any weight can be
employed in the construction of a Lie algebra, the weight 2 ones are
the most simple and natural choice. In the present work, we have not
addressed the problem of the global equivalence of the scalar
densities found to the usual hamiltonian constraint of general 
relativity. We have also not discussed the conditions for the
constraint combinations obtained to be well-defined. For example, it
has already been noticed by Brown and Kucha\v{r} that the hamiltonian
vector field associated with the new scalars vanishes on the
constraint surface. The criteria determining the answer to these 
questions may depend on the context in which 
the new scalars will be applied.

There are certainly other interesting aspects of these gravitational
constraint
combinations. They are related to matter time in canonical gravity and
to coordinate conditions\cite{Kuchar}. Possibly, they can also
come out of a canonical transformation of the geometric canonical
data that separates the true degrees of freedom (in the spirit of
\cite{KuTime}). The possibilities they open in the quantum theory are
certainly worth investigating and we will discuss them in future
work.

\section{Acknowledgements}

I would like to thank Chris Isham and Stephan Schreckenberg for many
helpful discussions, and this work has also benefited from the suggestions
of the referees. The author is partly supported by the
A~S~Onassis Foundation.



\newpage

\begin{thebibliography}{99}



\bibitem{ADM}  Arnowitt R, Deser S and Misner S W 1962 in
 Gravitation: An Introduction to Current Research, ed
 Witten  L (New York: Wiley)


\bibitem{Dirac} Dirac P A M  1964 Lectures on Quantum Mechanics
(New York: Yeshiva University)

\bibitem{KuDGR} Kucha\v{r} K V 1981
in { Quantum Gravity 2: A second Oxford symposium}, ed
Isham~C~J, Penrose R and Sciama D W  (Oxford: Clarendon Press)

\bibitem{Sund} Sundermeyer K 1982 Constrained Dynamics 
(Berlin: Springer--Verlag)

\bibitem{BrKu}  Brown J D and  Kucha\v r K V 1995 {\it Phys. Rev.\/} D
{\bf 51} 5600 
 
\bibitem{Kuchar} Kucha\v{r} K V and Torre C G 1991 {\it Phys. Rev.\/} D
{\bf 43} 419; D {\bf 44} 3116\newline
Stone C L and Kucha\v{r} K V 1991 {\it Class. Quantum Grav.\/} 
{\bf 9} 757\newline
Kucha\v{r} K V 1991 {\it Phys. Rev.\/} D {\bf 43} 3332; D {\bf
44} 43 

\bibitem{Brown}  Brown J D 1993 {\it Class. Quantum Grav.} {\bf 10}
1579\newline
 Brown J D 1994 ``On variational principles for gravitating
perfect fluids'' (gr-qc/9407008)

\bibitem{KuRo} Kucha\v{r} K V and  Romano J D 1995 {\it Phys. Rev.\/} 
{\bf 51} 5579 

\bibitem{iannis} Kouletsis I 1996  ``Action functionals of single scalar
 fields and arbitrary--weight gravitational constraints that generate 
 a genuine Lie algebra'' (gr-qc/9601043)

\bibitem{iannis2} Kouletsis I, private communication.

\bibitem{pde} Garabedian P R 1967 Partial Differential Equations
(New York: Wiley)

\bibitem{BrMa} DeWitt B S 1967 Phys. Rev. {\bf 160} 1113\newline
Brown J D and  Marolf D 1995 ``On
relativistic material reference frames'' (gr-qc/9509026)

\bibitem{KuTime}  Kucha\v{r} K V 1992  ``Matter time in canonical quantum
gravity'' University of Utah preprint UU-REL-92-12-10 



\end{thebibliography}
\end{document}